\documentclass[twocolumn,showpacs,preprintnumbers,amsmath,amssymb]{revtex4}

\usepackage{graphicx}
\usepackage{dcolumn}
\usepackage{bm}

\begin{document}

\preprint{PRESAT-7902}

\title{First-Principles Study on Electron-Conduction Properties of Helical Gold Nanowires}

\author{Tomoya Ono}
\affiliation{Research Center for Ultra-Precision Science and Technology, Osaka University, Suita, Osaka 565-0871, Japan}

\author{Kikuji Hirose}
\affiliation{
Department of Precision Science and Technology, Osaka University, Suita, Osaka 565-0871, Japan}

\date{\today}

\begin{abstract}
Multishell helical gold nanowires (HGNs) suspended between semi-infinite electrodes are found to exhibit peculiar electron-conduction properties by first-principles calculations based on the density functional theory. Our results that the numbers of conduction channels in the HGNs and their conductances are smaller than those expected from a single-atom-row nanowire verify the recent experiment. In addition, we obtained a more striking result that in the cases of thin HGNs, distinct magnetic fields are induced by the electron current helically flowing around the shells. This finding indicates that the HGNs can be good candidates for nanometer-scale solenoids.
\end{abstract}

\maketitle
As the miniaturization of electronic devices progresses, the fabrication of atomic-scale structures and the elucidation of their electronic properties have attracted considerable attention. Within the last decade, a large number of experiments concerning electron-conduction properties of nanowires of a few nanometer lengths have been carried out using a scanning tunneling microscope or a mechanically controllable break junction \cite{datta,ruitenbeek}. Among the most exciting discoveries are the tip-suspended multishell helical gold nanowires (HGNs) of ~1 nm diameter and ~5 nm length observed by Kondo and Takayanagi \cite{kondo} using transmission electron microscopy (TEM). The HGNs are sandwiched between Au(110) electrodes and composed of several atom rows twisting around their axis. In such minute structures, it has been well known that electron transport becomes ballistic and exciting quantum phenomena take place. For example, it was experimentally observed that as a gold point contact is stretched and thinned, the electronic conductance of the contact decreases roughly in units of the quantized value, G$_0=2e^2/h$, where $e$ is the electron charge and $h$ is Plank's constant \cite{rubio,yanson,ohnishi,kizuka}. When it comes to HGNs, since a few relationships between their geometries and conductances have been investigated using TEM images \cite{ohshima}, there remains much to be learned about their electronic characteristics; in particular, the electron-conduction properties of these structures are important from both fundamental and practical points of view. Recently, Tosatti {\it et al.} \cite{tosatti} explored the geometries and electronic states of infinitely periodic helical wires using conventional first-principles calculations with a plane-wave basis set. However, the conduction properties of nanowires intervening between electrodes cannot be determined from the results of infinitely periodic wires because scatterings of electrons, which play crucial roles in the electron conduction, do not take place in the infinitely periodic systems. Therefore, in order to interpret the conduction properties profoundly by theoretical calculations, the models in which the nanowires are connected to semi-infinite electrodes are indispensable.

In this study, we examined the electron-conduction properties of HGNs suspended between semi-infinite gold electrodes using first-principles calculations within the framework of the density functional theory. Our findings are that the numbers of conduction channels of the HGNs do not coincide with the numbers of atom rows in the HGNs, while the number of conduction channels in a single-atom-row nanowire is one. In addition, some channels of the HGNs are not fully open, whereas the channel transmission of the single-atom-row nanowire is close to one. As a consequence, the conductances of the HGNs are not quantized and become much smaller than those expected from a single-atom-row nanowire. More intrigingly, in some of the HGNs, electron currents rotate around the nanowire axis so as to induce a magnetic field.

Our first-principles calculation method for electron-conduction properties is based on the real-space finite-difference approach \cite{chelikowsky}, which enables us to determine the self-consistent electronic ground state with a high degree of accuracy by the timesaving double-grid technique \cite{tsdg} and the direct minimization of the energy functional \cite{dm}. Moreover, the real-space calculations eliminate the serious drawbacks of the conventional plane-wave approach such as its inability to describe strictly nonperiodic systems. The local pseudopotential of a gold 6s electron constructed using the algorithm by Bachelet {\it et al.} \cite{hsc} is employed to describe the ion-core potential, and exchange correlation effects are treated by the local density approximation \cite{lda}. We take a cutoff energy of 12 Ry, which corresponds to a grid spacing of 0.91 a.u., and a higher cutoff energy of 110 Ry in the vicinity of nuclei with the augmentation of double-grid points.

\begin{table}[h]
\begin{center}
\caption{Numbers of atoms in the each shell of the nanowire.}
\label{tbl:table1}
\begin{tabular}{c|ccc} \hline\hline
& Innermost & \hspace{2mm}2nd\hspace{2mm} & \hspace{2mm}3rd\hspace{2mm} \\ \hline
Single-row &   9  & ---  & ---  \\
7-1        &   9  &  63  & ---  \\
11-4       &  24  & 101  & ---  \\
13-6       &  54  & 121  & ---  \\
14-7-1     &   9  &  63  & 133  \\
15-8-1     &   9  &  76  & 139  \\ \hline\hline
\end{tabular}
\\
\end{center}
\end{table}

\begin{table*}[tbp]
\caption{Conductances and channel transmissions at the Fermi level}
{\tabcolsep=2.5mm
\begin{tabular*}{110mm}{ccccccc}
\hline \hline
 & Single-row & 7-1 & 11-4 & 13-6 & 14-7-1 & 15-8-1 \\ \hline
Conductance (G$_0$) & 0.96  & 5.19   & 9.08   & 11.97  & 13.82  & 14.44 \\ \hline
1st ch.     & 0.958 & 0.997  & 1.000  & 0.996  & 1.000  & 1.000 \\
2nd ch.     & ---   & 0.995  & 0.991  & 0.992  & 0.999  & 0.996 \\
3rd ch.     & ---   & 0.970  & 0.986  & 0.982  & 0.991  & 0.991 \\
4th ch.     & ---   & 0.938  & 0.890  & 0.978  & 0.984  & 0.982 \\
5th ch.     & ---   & 0.653  & 0.884  & 0.962  & 0.959  & 0.975 \\
6th ch.     & ---   & 0.640  & 0.874  & 0.958  & 0.947  & 0.954 \\
7th ch.     & ---   & ---    & 0.753  & 0.940  & 0.940  & 0.942 \\
8th ch.     & ---   & ---    & 0.748  & 0.878  & 0.927  & 0.931 \\
9th ch.     & ---   & ---    & 0.620  & 0.872  & 0.893  & 0.899 \\
10th ch.    & ---   & ---    & 0.517  & 0.744  & 0.888  & 0.891 \\
11th ch.    & ---   & ---    & 0.497  & 0.735  & 0.881  & 0.841 \\
12th ch.    & ---   & ---    & 0.318  & 0.706  & 0.867  & 0.834 \\
13th ch.    & ---   & ---    & ---    & 0.618  & 0.770  & 0.732 \\
14th ch.    & ---   & ---    & ---    & 0.383  & 0.700  & 0.691 \\
15th ch.    & ---   & ---    & ---    & 0.200  & 0.542  & 0.645 \\
16th ch.    & ---   & ---    & ---    & 0.002  & 0.527  & 0.595 \\
17th ch.    & ---   & ---    & ---    & ---    & ---    & 0.537 \\ \hline \hline
\end{tabular*}}
\label{tbl:table2}
\end{table*}

\begin{figure*}[htb]
\begin{center}
\vspace{9.03cm}
\end{center}
\caption{(color) Isosurfaces of channel electron distributions of single-atom-row and 7-1 nanowires. (a) The first channel of the single-atom-row nanowire, (b) the first channel, (c) the third channel, and (d) the fourth channel of the 7-1 nanowire. The states incident from the left electrode are shown. The yellow and green spheres represent the atoms in the outer and inner shells, respectively. The rectangular parallelepipeds at both the edges of the nanowires are jellium electrodes.}
\label{fig:fig1}
\end{figure*}

The computational models are as follows: The HGNs are sandwiched between semi-infinite gold electrodes imitated by the structureless jellium $r_s$=3.01 (see Fig.~\ref{fig:fig1}, for example.). A unit cell of 40.1 a.u. in the $x$ and $y$ directions with periodic boundary conditions is adopted, where the $x$ and $y$ directions are perpendicular to the nanowire axis. The distance between jellium electrode surfaces is fixed at 51.2 a.u. The atomic geometries of the HGNs are determined using the parameters evaluated by Kondo and Takayanagi \cite{kondo}, and their lengths are set at $\sim$ 49.1 a.u. Table~\ref{tbl:table1} shows the number of atoms in the HGNs. The global wave functions for infinitely extended states from one electrode side to the other are evaluated by the overbridging boundary-matching formula \cite{fujimoto1,lang,fujimoto2}. The conductance of the nanowire system in the limits of zero temperature and zero bias is described by the Landauer-B\"uttiker formula G=tr({\bf T}$^\dag${\bf T}) G$_0$ \cite{buttiker}, where {\bf T} is the transmission matrix. The eigenchannels are computed by diagonalizing the Hermitian matrix {\bf T}$^\dag${\bf T} \cite{nkobayashi}.

\begin{figure}[htb]
\begin{center}
\vspace{14.65cm}
\end{center}
\caption{(color) Induced magnetic flux density per bias voltage. (a) 7-1 nanowire, (b) 11-4, and (c) 15-8-1 nanowire. The gray, yellow and green spheres represent the atoms in the innermost, second, and third shells, respectively. The broken lines are the edges of the jellium electrodes.}
\label{fig:fig2}
\end{figure}

Table~\ref{tbl:table2} lists the conductances and the channel transmissions of the single-atom-row, 7-1, 11-4, 13-6, 14-7-1, and 15-8-1 nanowires suspended between gold electrodes. In the case of the single-atom-row nanowire, the number of conduction channels is one and the conductance is quantized as $\sim$ 1 G$_0$, which are in good agreement with the results of previous theoretical calculations and experiments \cite{rubio,yanson,ohnishi,fujimoto1,okamoto1,smit}. In the case of a thicker HGN composed of several atom rows, the number of conduction channels does not correspond to the number of atom rows, and some channels are not fully open. The reduction of the number of conduction channels is mainly due to the geometry and electronic states of a nanowire rather than the interaction between the nanowire and electrodes; in fact, the number of channels in the 7-1 nanowire agrees with the number of energy bands crossing the Fermi level in the infinitely periodic wire \cite{tosatti}. On the other hand, the channel transmissions are closely related to the interfaces between the nanowire and electrodes; when we replaced the jellium electrodes with Au(110) crystalline electrodes, the conductance of the 7-1 nanowire becomes 4.8 G$_0$, whereas the number of conduction channels remains six. Consequently, the conductance per atom row of the HGNs is much smaller than that of the single-atom-row nanowire. Compared with recent experimental results \cite{ohshima}, the calculated conductances agree with those of the experiments within the relative difference of ~20 \%. These results imply that the electron-conduction channels of the HGNs are different from that of the single-atom-row nanowire; as an example, we show in Fig.~\ref{fig:fig1} some of the channel electron distributions of 7-1 nanowire, in which the states incident from the left electrode are depicted. For comparison, that of the single-atom-row nanowire is also described [Fig.~\ref{fig:fig1} (a)]. One can easily recognize that the conduction channels do not completely form along the atom rows in the 7-1 nanowire, while the conduction channel of the single-atom-row nanowire follows the atom row. Moreover, in Fig.~\ref{fig:fig1} (d), the fourth channel of the 7-1 nanowire shapes on the outer shell and twists about the nanowire axis with a right-hand rotation, which is opposite to that of the atom rows. The helical channel current is mainly caused by the asymmetry of the atomic structure of the HGN.

We next investigate the magnetic field induced by the electron current flowing through the HGNs. Since some channels twist around the nanowire axis, the helical electronic current through the nanowires gives rise to the magnetic field inside the nanowire (see Fig.~\ref{fig:fig2}). The direction of the induced magnetic field approximately agrees with that expected from the channel electron distribution. The magnetic flux density per bias voltage (10$^{-3}$ T/mV) is of the same order as those obtained using the channel electron distribution and its transmission. It is worth noting that there is a considerable difference in the strength of the magnetic field between the thin and thick HGNs: the field in the thin nanowire is stronger than that in the thick one. In the case of the thick HGN, the axial symmetry of its atomic structure breaks very moderately, and then no distinct magnetic field is induced because the electronic current does not twist largely around the nanowire axis [see Fig.~\ref{fig:fig2} (c)]. Okamoto {\it et al.} \cite{okamoto2,okamoto3} attempted to explore the magnetic field using nanowires made of structureless solid jellium and reported that the strength of the magnetic field is very small. In the case of the jellium model, the shell structures of the HGNs are not taken into account, therefore the helical currents flowing around the shells are not observed; the shell structures play important roles in the induction of magnetic fields by the helical currents.

To the best of our knowledge, the findings regarding the electron-conduction behavior (conductance, conduction-electron distribution, and so on) of the HGNs and the magnetic field induced by the helical current are the first theoretical indications, and experimental consequences should be interesting to pursue. Although whether the magnetic field would be measurable in steady states similarly to the case of macroscopic solenoids remains an unsettled question, our findings indicate that the HGNs are good candidates for nanometer-scale solenoids. Thus, this situation will stimulate a new technology for electronic devices using atomic wires. The work in progress is a simulation of the formation of the magnetic field by the time-dependent density functional theory.

This research was supported by a Grant-in-Aid for the 21st Century COE ``Center for Atomistic Fabrication Technology'' and also by Young Scientists (B) (Grant No. 14750022) from the Ministry of Education, Culture, Sports, Science and Technology. The numerical calculation was carried out by the computer facilities at the Institute for Solid State Physics at the University of Tokyo, and the Information Synergy Center at Tohoku University.

\end{document}